# Design and numerical investigation of cadmium telluride (CdTe) and iron silicide ($FeSi_2$) based double absorber solar cells to enhance power conversion efficiency

Md. Ferdous Rahman[1,4*], M. J. A. Habib[1], Md. Hasan Ali[1], M. H. K. Rubel[2], M. Rounakul Islam[3], Abu Bakar Md. Ismail[4], M. Khalid Hossain[5,**]

[1]*Department of Electrical and Electronic Engineering, Begum Rokeya University, Rangpur 5400, Bangladesh*
[2]*Department of Materials Science and Engineering, University of Rajshahi, Rajshahi 6205, Bangladesh*
[3]*Department of Electrical and Electronic Engineering, Southeast University, Dhaka 1208, Bangladesh*
[4]*Solar Energy Laboratory, Department of Electrical and Electronic Engineering, University of Rajshahi, Rajshahi 6205, Bangladesh*
[5]*Institute of Electronics, Atomic Energy Research Establishment, Bangladesh Atomic Energy Commission, Dhaka 1349, Bangladesh*

Correspondence: *ferdous@brur.ac.bd (***M.F. Rahman***), and **khalid.baec@gmail.com, (***M.K. Hossain***)

**Abstract**

Inorganic CdTe and $FeSi_2$-based solar cells have recently drawn a lot of attention because they offer superior thermal stability and good optoelectronic properties compared to conventional solar cells. In this work, a unique alternative technique is presented by using $FeSi_2$ as a secondary absorber layer and $In_2S_3$ as the window layer for improving photovoltaic (PV) performance parameters. Simulating on SCAPS-1D, the proposed double-absorber (Cu/FTO/$In_2S_3$/CdTe/$FeSi_2$/Ni) structure is thoroughly examined and analyzed. The window layer thickness, absorber layer thickness, acceptor density ($N_A$), donor density ($N_D$), defect density ($N_t$), series resistance ($R_S$), and shunt resistance ($R_{sh}$) were simulated in detail for optimization of the above configuration to improve PV performance. According to this study, 0.5 μm is the optimized thickness for both the CdTe and $FeSi_2$ absorber layers in order to maximize efficiency ($\eta$). Here, the value of the optimum window layer thickness is 50 nm. For using CdTe as a single absorber, the $\eta$ is achieved by 13.26%. But for using CdTe and $FeSi_2$ as a dual absorber, the $\eta$ is enhanced and the obtaining value is 27.35%. The other parameters are also improved and the obtaining values for fill factor (*FF*) are 83.68%, open-circuit voltage ($V_{oc}$) is 0.6566V, and short circuit current density ($J_{Sc}$) is 49.78 mA/cm$^2$. Furthermore, the proposed model performs good at 300 K operating temperature. The addition of the $FeSi_2$ layer to the cell structure has resulted in a significant quantum efficiency (*QE*) enhancement because of the rise in solar spectrum absorption at longer wavelengths ($\lambda$). The findings of this work offer a promising approach for producing high-performance and reasonably priced CdTe-based solar cells.

**Keywords:** Solar cell; double absorber layer; $FeSi_2$; $In_2S_3$; SCAPS-1D.

## List of abbreviations

| | | | |
|---|---|---|---|
| $\mu_h$ | hole mobility | $N_D$ | donor density |
| $\mu_n$ | electron mobility | **Ni** | nickel |
| **CdTe** | cadmium telluride | $N_{it}$ | interface defect |
| **CIGS** | copper indium gallium sulphide | $N_t$ | defect density |
| **Cu** | copper | ***PCE*** | power conversion efficiency |
| **DOS** | density of states | **PV** | photovoltaic |
| **DSSC** | dye-sensitized solar cells | ***QE*** | quantum efficiency |
| $E_g$ | bandgap | $R_S$ | series resistance |
| ***EQE*** | external quantum efficiency | $R_{sh}$ | shunt resistance |
| ***E*** | experimental | **SCAPS-1D** | one dimensional solar cell capacitance simulator |
| $FeSi_2$ | iron silicide | **Si** | silicon |
| ***FF*** | fill factor | **TFSC** | thin film solar cell |
| **FTO** | fluorine-doped tin oxide | ***T*** | theoretical |
| $In_2S_3$ | indium sulphide | $V_{OC}$ | open-circuit voltage |
| $J_{SC}$ | short-circuit current density | $\alpha$ | absorption coefficient |
| ***J-V*** | current voltage density | $\lambda$ | wavelength |
| $\eta$ | efficiency | $\chi$ | electron affinity |
| $N_A$ | acceptor density | | |

## 1 Introduction

The problem of global warming has increased scientific research into solar energy and other new energy sources. Solar cells are one of the best methods to produce electric power using solar energy [1–4]. Solar cells are constructed from a number of materials, including silicon (Si), which is the most commercially feasible and typical material [5–7]. Alternative materials have been analyzed with the majority of objectives in mind for developing solar cells that are low-cost, high-efficiency, and long-lasting [8–10]. Solid-state system encourages a large number of p-n junction solar PV devices and efforts to minimize existing dependency on coal and gas, due to this, the lower detrimental release of greenhouse gases [11–18].

In recent years, thin film solar cell (TFSC) technology including dye-sensitized solar cells (DSSCs) has gotten a lot of interest in the research community [19–22]. But due to the low efficiency and poor stability of DSSCs, the copper indium gallium sulphide (CIGS) and CdTe solar cells with a higher than 20% efficiency are attracted much [23]. For having some exclusive features, the thin film of CdTe is employed largely. Due to its abundance, lifetime stability, remarkable efficiency, strong optical absorption, and low cost, CdTe thin-film solar cells are gaining popularity in both electrical and optical applications. It is used in various devices such as solar cells, nanodevices, and sensors [24]. CdTe is a classification of an II-VI dichalcogenide transition metal with a high absorption coefficient ($\alpha$) (>$10^5$ $cm^{-1}$) and that is surely larger than other common semiconductor materials having a narrow bandgap (*Eg*) (~1.5 eV) [25]. For the solar light spectrum, the value of the $E_g$ is suitable [26]. Another interesting thing to consider is that CdTe has the advantage of being able to store both p-type and n-type conductivity. Since 1890, CdTe has been the most researched material; yet, it has only been exploited as a polycrystalline thin films and quantum dots material for the last ten years. For improving the stability and efficiency of CdTe different efforts have been planned [27]. Formation of a p-type semiconductor, the material CdTe is known as a potential absorber and is used for thin-film PV technology [28].

In some experimental works, for the widespread use of CdTe as a thin film, it has an early form of low power conversion efficiency in the power conversion efficiency (*PCE)* method [29,30]. The conversion efficiency of CdTe-based thin-film solar cells is 17.5% or additional in certain situations [26]. A research for the published (ZnO / CdS / CdTe / ZnTe) structure is reported to 24.66% with ($V_{oc}$ = 946.51 mV, $J_{sc}$ = 34.40 mA $cm^{-2}$ and $FF$ = 75.72%) under the global AM1.5G illumination spectrum [31]. Another better $\eta$ 26.3% is noticed with $J_{SC}$ (21.4 mA $cm^{-2}$), $V_{OC}$ (1.63 V), and $FF$ (74.86%) has been reported on an improved efficient perovskite / $FeSi_2$ monolithic tandem solar cell [32]. In another experimental work, the maximum efficiency of 24.35% is also noticed for structure (FTO/ZMO/$CdSe_xTe_{1-x}$/CdTe) Where $CdSe_xTe_{1-x}$ is used as a second absorber [33].

In this study, for increasing the low *PCE* performance, the structure (Cu/$In_2S_3$/CdTe/$FeSi_2$/Ni) is used for the configuration of the CdTe and $FeSi_2$-based solar cell device. In this configuration, non-toxic compounds such as $FeSi_2$ and $In_2S_3$ are combined with CdTe as well as various components. The absorber layer $FeSi_2$ has a direct band gap energy ($E_g$) of 0.87 eV [1] and a high $\alpha$ exceeding $10^5$ $cm^{-1}$ in photon energy of 1 eV [34–36]. In the Earth's crust, the component member elements Iron (Fe) and Si of $FeSi_2$ are the most plentiful elements. As a result, it has become a low-cost component that may be used to make active absorber materials for solar PV cells. Indium sulfide ($In_2S_3$) is a larger $E_g$ (~ 2.82ev) material and also has a heavy electron affinity ($\chi$) (~ 4.5 eV) [37]. It is used as the window layer in this study. The properties of electrical, optical, and structural $In_2S_3$ films are effective in optoelectronic devices, especially in various scientific, technological, and commercial applications which are engaging for solar cells [2]. In this study, the thickness of the absorber and window layers, the defect and acceptor densities of the absorber layer, and the $R_s$ and $R_{sh}$ of the full cell structure were optimized to maximize the *PCE*.

## 2 Modeling and Simulation

The structure of the proposed device for this simulated work investigation is illustrated in **Figure 1(a)**. In this research, the structure (Cu/FTO/$In_2S_3$/CdTe/$FeSi_2$/Ni) is the CdTe and $FeSi_2$- based solar cell with a p-type (CdTe) absorber having a higher $E_g$ of 1.5 eV [38] and $p^+$ -type another absorber ($FeSi_2$) layer having a lower $E_g$ of 0.875 eV, making a junction between them, while the value of $E_g$ of window layer ($In_2S_3$) to be 2.82 eV [37]. The proposed double absorber CdTe & $FeSi_2$-based configuration is presented to enhance the performance. The second absorber layer ($FeSi_2$) is put between the CdTe layer and the Nickel (Ni) back contact. After all the traditional $SnO_2$ has an $E_g$ of 3.6 eV which is next to the window layer and performs as an FTO layer [39–42]. This innovative solar cell design features a 0.5-$\mu m$ p-type absorber (CdTe) with $N_A$ =$1\times10^{10}$ $cm^{-3}$, a 0.5-µm p+-type additional absorber ($FeSi_2$) with $N_A$ =$1\times10^{17}$ $cm^{-3}$ and on the back contact Ni layer, a 0.05$\mu m$ n-type window layer ($In_2S_3$), and a 0.04 $\mu m$ as FTO ($SnO_2$) layer. Here, the front grid contact is made of Copper (Cu), which has a work function (WF) of 4.65 eV [43], and the back contact is Ni which has a work function of 5.25 eV [43].

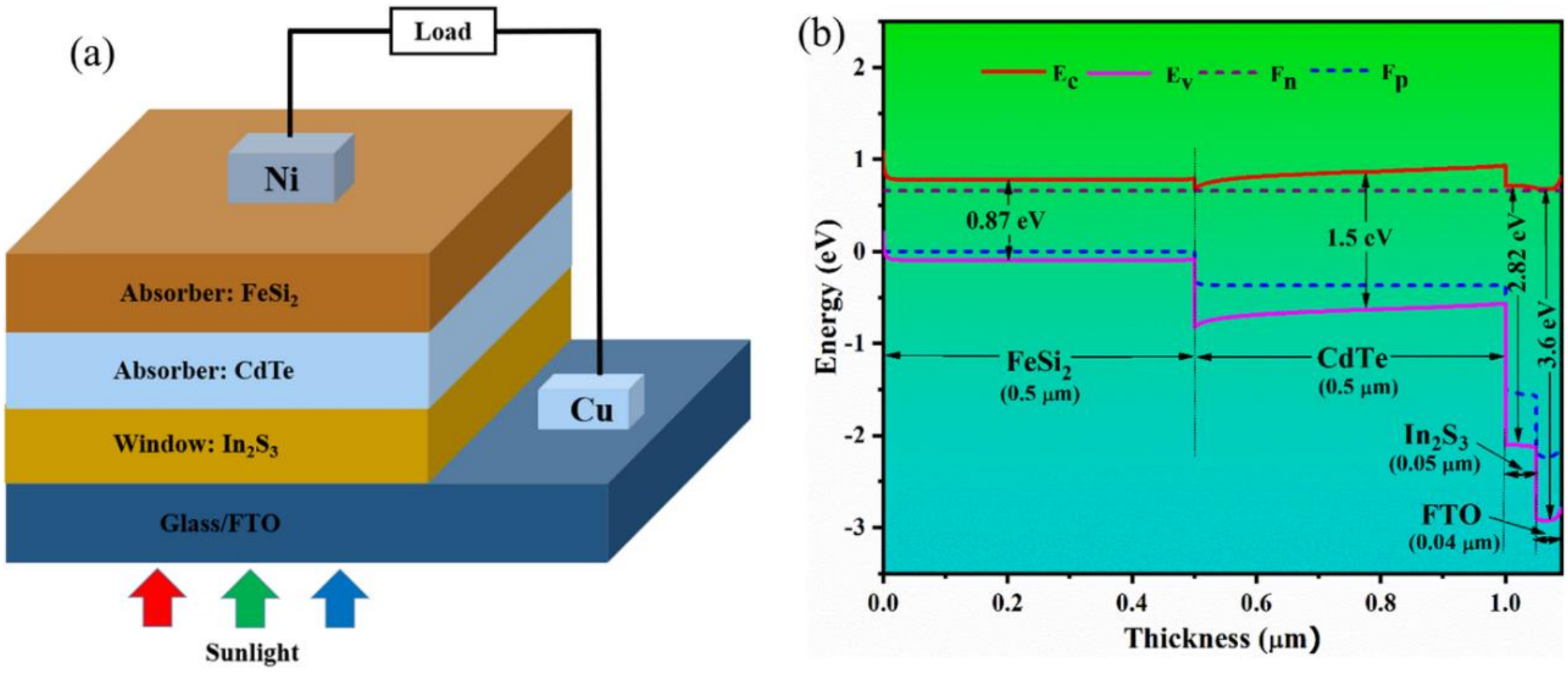

**Figure 1.** The CdTe and $FeSi_2$-based solar cell: (a) proposed structure, (b) energy band diagram.

Researchers use software packages including AMPS [44], SILVACO ATLAS [45], COMSOL [46], wxAMPS [47], and SCAPS [48] to investigate TFSCs performance. In the current study, we analyzed the performance of solar cells and their various determining characteristics using the SCAPS one-dimensional solar cell simulator (version 3.3.07). The SCAPS software provides many benefits, including the capacity to examine performance at seven various levels, including numerous depth and batch analyses, as well as the ability to swiftly comprehend and analyze data [49].

In addition, the results gained with the SCAPS simulator are compatible with other researchers' earlier experimental results. Several research papers on the SCAPS-1D program have recently been published, examining its use in determining the efficiencies of various kinds of solar cells [50,51]. Using three equations of semiconductor physics, SCAPS is based on which are most important. These are the continuity equations for free electrons and holes and Poisson's equation. Almost all the parameters, including $E_g$, doping density, $\chi$, the effective density of states (DOS), mobility, etc. can be graded in SCAPS. Varieties of illumination spectrums, such as AM0, AM1.5G, monochromatic, white, and so on, are available with standard test conditions (STC) for illumination spectrum (Global Air Mass, AM 1.5G).

The values in **Table 1** represent the parameters of the layers that were used as input to the simulation. Band alignment determines how much current flows via the heterojunction. **Figure 1(b)** represents the $E_g$ and layer thickness of each component based on the energy band panel data received after the SCAPS simulation for this investigation. According to **Figure 1(b)**, it is possible to identify easily the band bending that happens between $FeSi_2$ and CdTe junction because of using the various doping concentration levels in this study.

**Table 1.** Taking parameters for (Cu/FTO/$In_2S_3$/CdTe/$FeSi_2$/Ni) solar cell. Front contact: Cu [43], FTO as $SnO_2$:F [19,39,41,42,52–54], $In_2S_3$ as window layer [37,41,55], absorber layer as CdTe [19,38–40,42], absorber layer as $FeSi_2$ [1,34–36,40,41], back contact: Ni [43].

| **Parameter (unit)** | **n-type window ($In_2S_3$)** | **FTO** | **p-type absorber CdTe** | **$p^+$-type absorber $FeSi_2$** |
|---|---|---|---|---|
| Thickness (μm) | 0.05 | 0.04 | 0.5 | 0.5 |
| $E_g$ (eV) | 2.82 | 3.6 | 1.5 | 0.87 |
| Electron affinity (eV) | 4.5 | 4.5 | 4.28 | 4.16 |
| Dielectric permittivity (relative) | 13.5 | 10 | 10.30 | 22.5 |
| CB effective DOS ($cm^{-3}$) | $2.2 \times 10^{17}$ | $2.00 \times 10^{18}$ | $9.2\times 10^{17}$ | $5.6\times 10^{19}$ |
| VB effective DOS ($cm^{-3}$) | $1.8 \times 10^{19}$ | $1.8 \times 10^{19}$ | $5.2 \times 10^{18}$ | $2.08 \times 10^{19}$ |
| Electron thermal velocity (cm/s) | $1\times10^7$ | $2\times10^7$ | $1\times10^7$ | $1\times10^7$ |
| Hole thermal velocity (cm/s) | $1\times10^7$ | $1\times10^7$ | $1\times10^7$ | $1\times10^7$ |
| Electron mobility, $\mu_n$ ($cm^2$/V s) | 100 | 100 | $3.2\times10^2$ | 100 |
| Hole mobility, $\mu_h$ ($cm^2$/V s) | 25 | 20 | 40 | 20 |
| $N_D$ ($cm^{-3}$) | $1\times10^{13}$ | $1 \times 10^{18}$ | 0 | 0 |
| $N_A$ ($cm^{-3}$) | 0 | 0 | $1 \times 10^{10}$ | $1 \times 10^{17}$ |
| Defect type | Neutral | Neutral | Neutral | Neutral |

## 3 Results and discussion

The main goal of this research is to look into the impacts of various absorber layer parameters on solar cell photo conversion efficiency, as well as compare the effects of the same factors for CdTe and $FeSi_2$-based thin-film solar cells. With the aid of optimized data, we will be able to establish a set of standards for designing the most effective real-time solar PV devices.

The modeling and simulation studies were carried out in some stages: optimization of the (a) thickness of the CdTe and $FeSi_2$ layer, (b) thickness of the $In_2S_3$ window layer, (c) defect and carrier densities of the absorber layer, (d) $R_s$ and $R_{sh}$, and (e) work function of the back contact of the device.

Following optimization of the mentioned parameters, current-voltage density (*J-V*) characteristic analysis, carrier generation, and recombination analysis were performed on the optimized data to examine the junction and other device attributes. This in-depth investigation allows us to identify the shortcomings in CdTe and $FeSi_2$-based solar cells, allowing the scientific community to develop more efficient solar cell devices.

### 3.1 Impact of variation of CdTe Absorber Layer Thickness

#### *3.1.1 Case I. Conventional CdTe model (single absorber)*

The thickness of the absorber layer is a significant component in solar cells' performance. Comparing the findings to those from experimental fabrication [29–33] allows for the evaluation of CdTe cells. The impact of the thickness of the p-type CdTe absorber on the PV performance of the cell is shown in **Figure 2**, where the absorber layer thickness varies from 0.1 to 4.5 µm. Here under 1-sun illumination, 300 K temperature is used to apply the parameter values for the various layers as listed in **Table 1**. When the thickness of the CdTe absorber layer is raised, a significant number of photons from solar light is absorbed, improving $\eta$. The $\eta$ is 6.21 % and 15.44% for absorbers with thicknesses of 0.1 µm and 4.5 µm, respectively. **Figure 2** shows, how the $\eta$ rises as increasing the thickness. Additionally, $V_{OC}$ and $J_{SC}$ also changed in proportion to the absorber layer's thickness. *QE* is a measure of how much current a solar cell can generate when exposed to photons of a specific $\lambda$. The number of electrons that can be extracted from a PV device for every photon that is incident is measured by external quantum efficiency (*EQE*). The *QE* for the optimum thickness of the basic CdTe cell is shown in **Figure 3**.

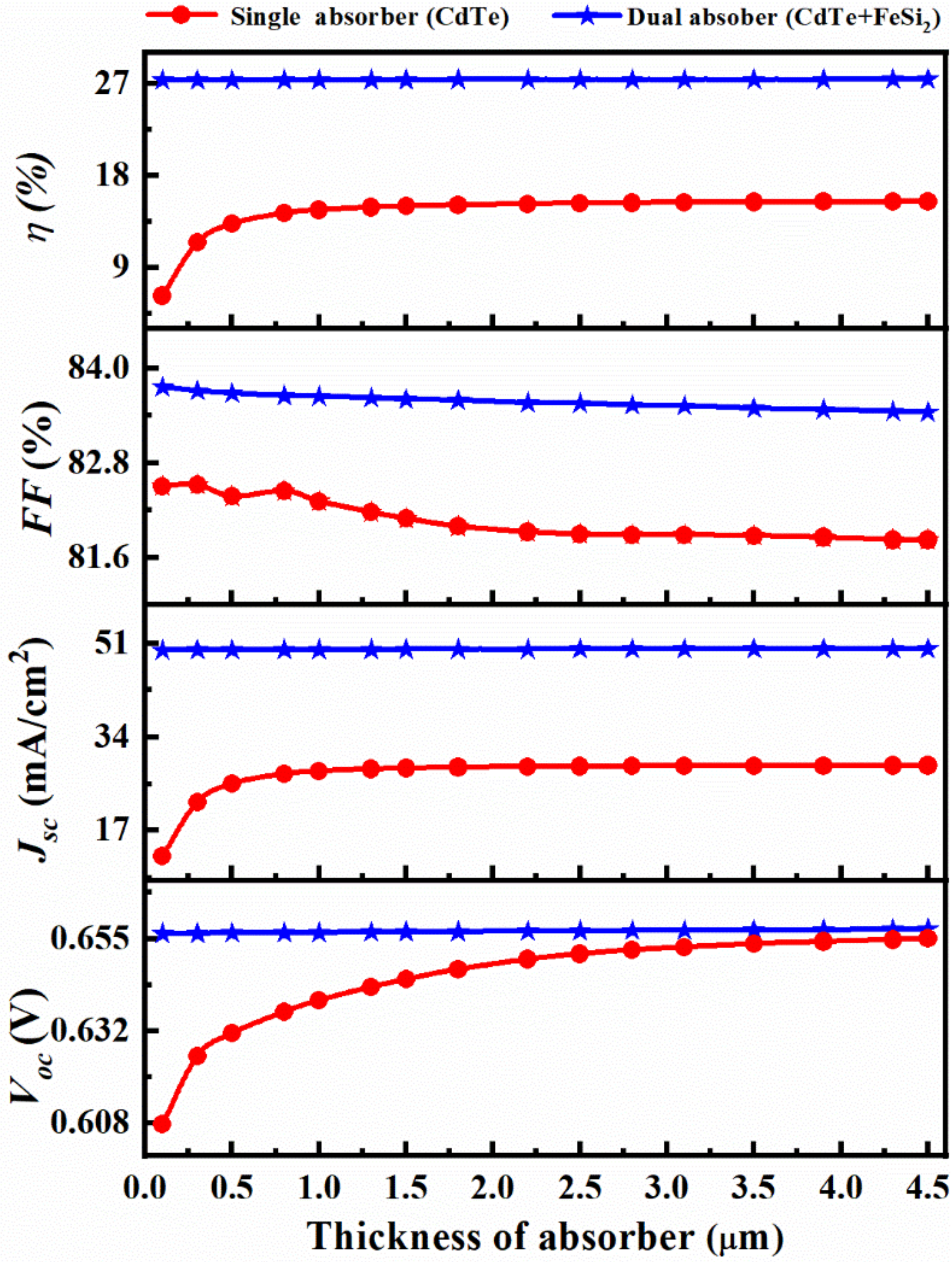

**Figure 2.** Effects of adding and removing extra absorber $FeSi_2$ when adjusting CdTe thickness.

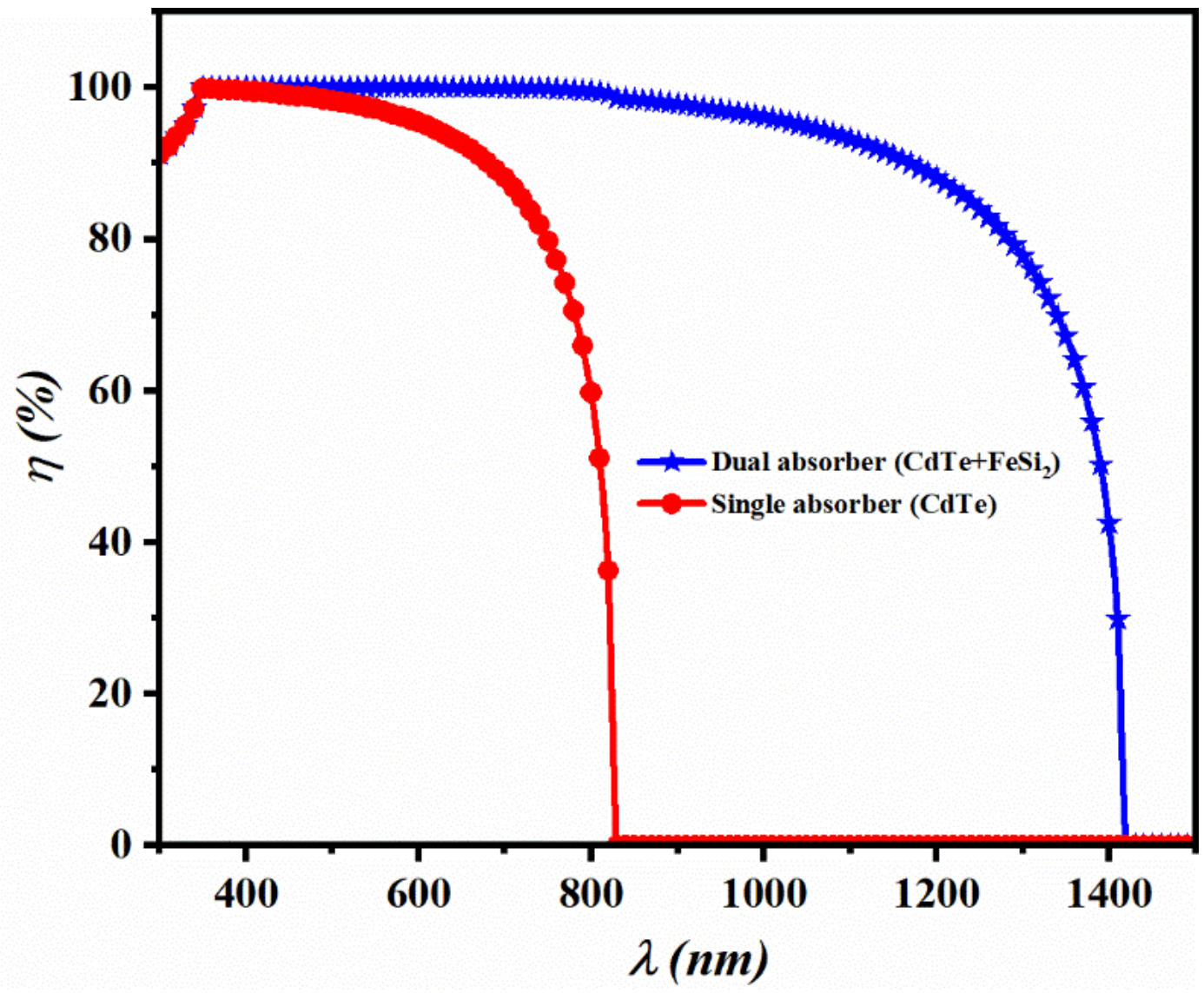

**Figure 3.** The variation *EQE* with $\lambda$ for the CdTe and $FeSi_2$ solar cell.

### 3.1.2 *Case II: CdTe solar cell with second absorber $FeSi_2$*

When an extra layer is added or the thickness of the absorber is increased, a high impurity doping concentration can avoid the significant recombination of the minority charge carriers (electrons) at the metallic back contact layer. But increasing thickness is not possible all the time because of the abundance of the desired material and it is also expensive. For this reason, in this study $FeSi_2$ (with doping of $1\times10^{17}$) is used adjacent to CdTe (with doping of $1\times10^{10}$). The interface between $p^+$-type $FeSi_2$ and p-type CdTe generates an electric field that inhibits the movement of minority carriers to the rear surface, just like in a p–n junction. Therefore, the $FeSi_2$ layer will boost short-circuit current while decreasing dark current by reflecting the minority carriers. The $\eta$ of the solar cell is increased by the inclusion of a second absorber layer by lowering the rate of surface recombination. At a temperature of 300 K and 1 sun-illumination, the parameter values listed in Table 1 are employed in this study. The thickness of the $FeSi_2$ layer is first varied from 0.1 µm to 4.5 µm. The cell's $\eta$ stabilizes between 0.3 µm to 1.8 µm, according to this initial investigation. The $\eta$ of the typical CdTe solar cell with a 0.1 µm absorber layer is 24.66%. Therefore, it is decided that a 0.5 µm thin $FeSi_2$ layer is appropriate and is utilized in the next studies. The redesigned CdTe solar cell demonstrates a 27.35% $\eta$ for the lower thickness of the CdTe layer of 0.5 µm, indicating a considerable increase in $\eta$ for the lower thickness of the absorber layer. This result is in good agreement with simulated works as well as experimental research in literature [29,32,33]. The primary cause of this is that the solar cell's $FeSi_2$ and CdTe layers both serve as absorbers, with a combined thickness of 1 µm. The $FeSi_2$ layer aids in the solar cell's ability to absorb more photon energy. Thus, a huge number of photons can be collected, leading to the generation of additional electron-hole pairs. Thus, incorporation of the $FeSi_2$ layer, $J_{SC}$ increases from 25.505 to 49.78 mA/cm$^2$ while $V_{OC}$ also increases from 0.631 to 0.656 V.

The performance of typical CdTe-based solar cells is greatly improved by all of these characteristics, which raise solar cell efficiency. However, the *FF* decreases from 83.44 to 82.5%. This is such that when $V_{OC}$ and $J_{SC}$ both increase, $V_{MPP}$ and $J_{MPP}$ do not increase but instead decrease simultaneously. Thus, *FF* seems to decrease according to **Eq. (1)** [56]. However, the *FF* still offers a better value for solar cell operation.

$$FF = \frac{V_{MPP} \times J_{MPP}}{V_{OC} \times J_{SC}} \tag{1}$$

where $V_{MPP}$ and $J_{MPP}$ denote the voltage and current, respectively, at the maximum power point. The impact of the optimum thicknesses of the *QE* is shown in **Figure 3**. Due to the improved photon capture by thicker absorber layers, the *QE* of the solar cell initially increases when the CdTe layer thickness is increased. Because each substance can only absorb photons in a specific region of the visible light spectrum, all the curves begin to slope downward towards zero QE at specific $\lambda$ points. Due to the availability of Cd and Te, it costs more money to have a better outcome with higher CdTe thickness. Reducing the thickness of the CdTe material may also reduce the amount of Cd and Te required, which may ultimately lead to lower production costs for such devices. Additionally, compared to the previous design, these results demonstrate increased $\eta$ when using the new cost-effective solar cell structure. Compared to Cd and Te, $FeSi_2$ is more inexpensive and available. The results of this study might thus assist manufacturing companies in producing CdTe solar cells at a larger profit.

## 3.2 Impact of changing $In_2S_3$ Window Layer Thickness

The thickness of the $In_2S_3$ window layer varied from 0.01 µm to 0.35 µm where other parameters had been maintained. A 0.05 µm thickness difference was taken from the given range and based on this parameter such as $J_{sc}$, $\eta$, $V_{oc}$, and *FF* were collected by using SCAPS-1D simulation. The cell output parameters had been shown in **Figure 4** through the variation of $In_2S_3$ window layer thickness. The efficiency conversion results 27.35% along with $V_{oc}$= 0.6566 V, $J_{sc}$ = 49.78 mA/cm$^2$, *FF* = 83.68% had finally been gained for dual absorber (CdTe and $FeSi_2$) and the thickness of $FeSi_2$ is 500 nm, 50 nm thickness is used for window layer ($In_2S_3$), and 40 nm FTO ($SnO_2$:F) and without absorber layer ($FeSi_2$) the $\eta$ is measured 13.26%, $V_{oc}$ = 0.6312 V, $J_{sc}$ = 25.51 mA/cm$^2$, 82.37% *FF* when the thickness of absorber (CdTe) is 0.5 µm.

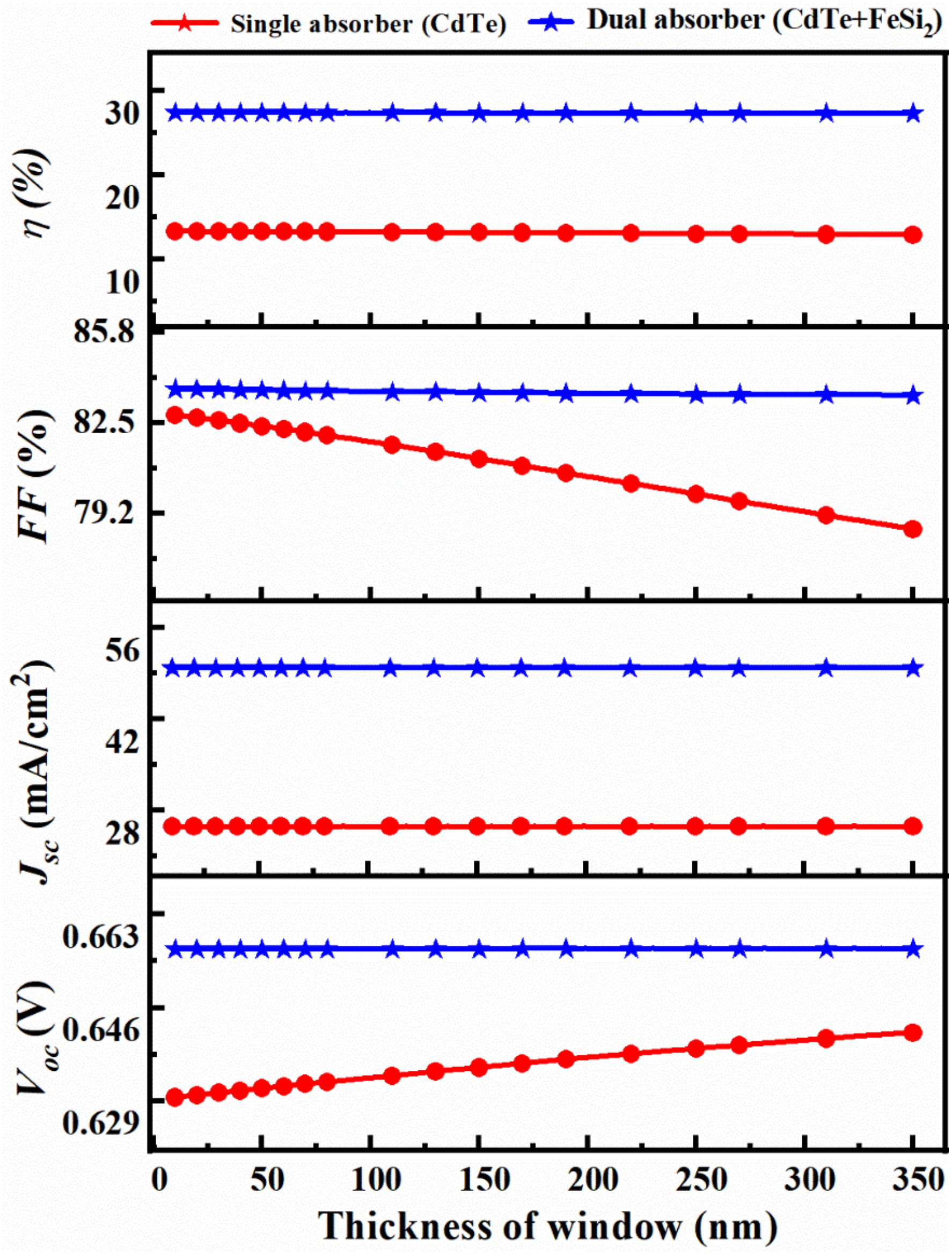

**Figure 4.** The effect of varying the $In_2S_3$ thickness with the single and dual absorber.

## 3.3 The effect of CdTe acceptor density on this solar cell's performance

$N_A$ in the absorber (CdTe) layer is varied from $1\times10^{10}$ to $1\times10^{19}$ cm$^{-3}$ for investigating its impact on the performance of the CdTe solar cells. For this case, the $N_A$ of the $FeSi_2$ is $1\times10^{17}$, and the $N_D$ of $In_2S_3$ and $SnO_2$:F is $1\times10^{13}$ and $1\times10^{18}$ respectively. Here **Figure 5(a)** displays the variation of the effect of the $N_A$ in the single absorber (CdTe) layer. The purpose of changing the $N_A$ value in the CdTe material from $10^{10}$ to $10^{19}$ resulted that increased $V_{oc}$ from 0.6566 to 0.9464 V, and $J_{Sc}$ reduces from 49.77 to 17.95 mA/cm$^2$. The *FF* remains constant up to $10^{15}$. The excess carrier causes recombination and increases scattering, which results in lower cell efficiency when the value of $N_A$ exceeds $1\times10^{15}$ cm$^{-3}$. Under Consideration of the low cost and growth rate of fabrication, the optimized condition of $N_A$ (CdTe) is found at the $10^{10}$ cm$^{-3}$.

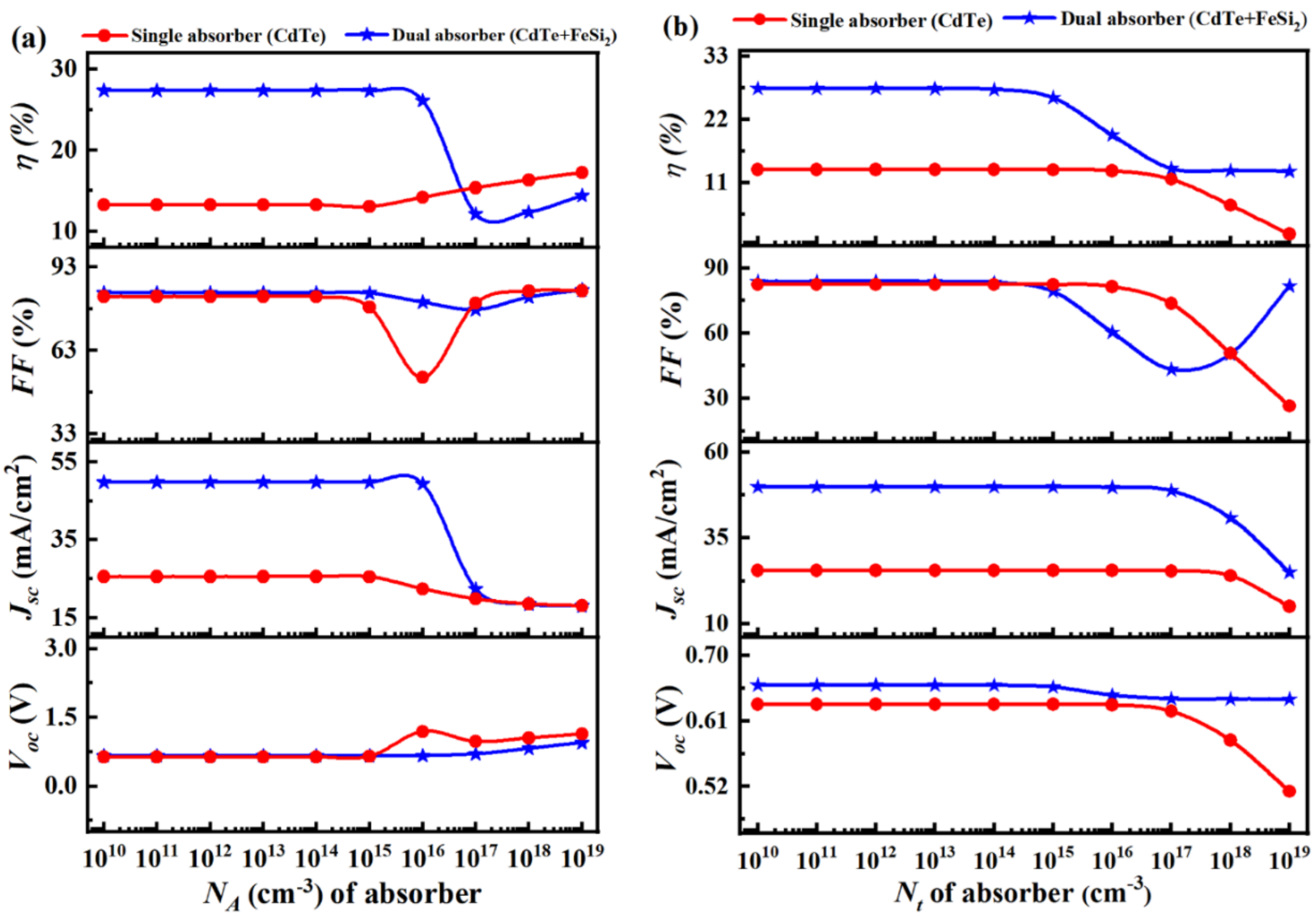

**Figure 5.** The effect of changing the (a) $N_A$, and (b) $N_t$ with the single and dual absorber.

### 3.4 Effect of defect density in the CdTe absorber layer

$N_t$ is another important factor that has a direct impact on the performance of solar cells. The absorber layer is primarily responsible for producing the photo-generated current. As a result, a rise in $N_t$ causes carrier recombination to increase, which lowers device $\eta$. This experiment varies the $N_t$ between $10^{10}$ cm$^{-3}$ to $10^{19}$ cm$^{-3}$ to examine the impact of the $N_t$. According to research, the $N_t$ significantly influences the solar cell's output parameter, as illustrated in **Figure 5(b)**. According to **Figure 5(b)**, the $\eta$ of CdTe and $FeSi_2$ solar cells starts to decrease when $N_t$ rise beyond $10^{14}$ cm$^{-3}$ and $10^{15}$ cm$^{-3}$, respectively. The values obtained during the simulation are in good agreement with the $N_t$ values. As shown in **Figure 5(b)**, the $V_{oc}$ and $J_{sc}$ for CdTe decrease with large values of $N_t$. According to **Figure 5(b)**, the *FF* exhibits the same pattern for solar cells made of CdTe and $FeSi_2$. For all levels of $N_t$, the *FF* for dual absorber is higher than for single absorber (CdTe). The carrier lifetime and diffusion length decrease as the $N_t$ and carrier recombination rate both rise. Consequently, the device's overall performance suffers. It is possible to lower the $N_t$ and increase $\eta$ by carefully adjusting the deposition parameters.

### 3.5 Effect of resistances on the suggested solar cell's PV performance

The performance of solar cell characteristics is significantly influenced by $R_s$ and $R_{sh}$. The junction properties and how they affect the functionality of the device are likewise determined by these factors. All other optimized parameters from the previous section were left fixed at their best values to assess the impact of $R_s$ and $R_{sh}$. The $R_s$ and $R_{sh}$ for this simulation were changed in the ranges of 0 to 7 Ω-cm$^2$ and $10^1$-$10^7$ Ω-cm$^2$, respectively. **Figure 6** illustrates how $R_s$ and $R_{sh}$ affect the factors that determine how a solar cell performs well. The effectiveness of solar cells is determined to be significantly influenced by the $R_s$ and $R_{sh}$, as shown in **Figure 6**. The $V_{oc}$ is more reliant on lower values of $R_{sh}$ than $R_s$, but the $J_{sc}$ is more sensitive to higher levels of $R_s$ than $R_{sh}$. Along with the change in $R_{sh}$, the *FF* also changes. The other parameters, such as absorber and window layer thickness, and $N_t$, were maintained at their ideal values as determined in the prior sections to analyze the effects of $R_s$ and $R_{sh}$. The ideal values for $R_s$ and $R_{sh}$ of CdTe and $FeSi_2$-based solar cell devices are found in this work to be between 0 and 2 Ω-cm$^2$ and $10^3$ and $10^5$ Ω-cm$^2$, respectively.

### 3.6 Impact of donor density in the $In_2S_3$ window layer

With 500 nm CdTe ($N_A$ = $1\times10^{10}$,) and 500 nm $FeSi_2$ thick absorber ($N_A$= $1 \times 10^{17}$,),50 nm thick $In_2S_3$ ($N_D$= $10^{13}$,) and 40 nm thick fluorine doped tin oxide (FTO) ($N_D$= $10^{18}$cm$^{-3}$), donor concentration in window layer, $N_D$ was varied from $1\times10^{10}$ cm$^{-3}$ to What value at 300 K as pictured in **Figure** 7(a). With increasing window layer $\boldsymbol{N_t}$, $V_{oc}$ remains constant. The value $V_{oc}$ is 0.6566 V. $J_{sc}$ is almost constant. *FF* increases slowly from putting *FF* value for $1\times10^{10}$ to $1\times10^{20}$ respectively $\eta$ starts increasing from $10^{17}$ cm$^{-3}$. Hence considering overall performance, $N_D$ of $1\times10^{13}$ cm$^{-3}$ for the window layer has opted.

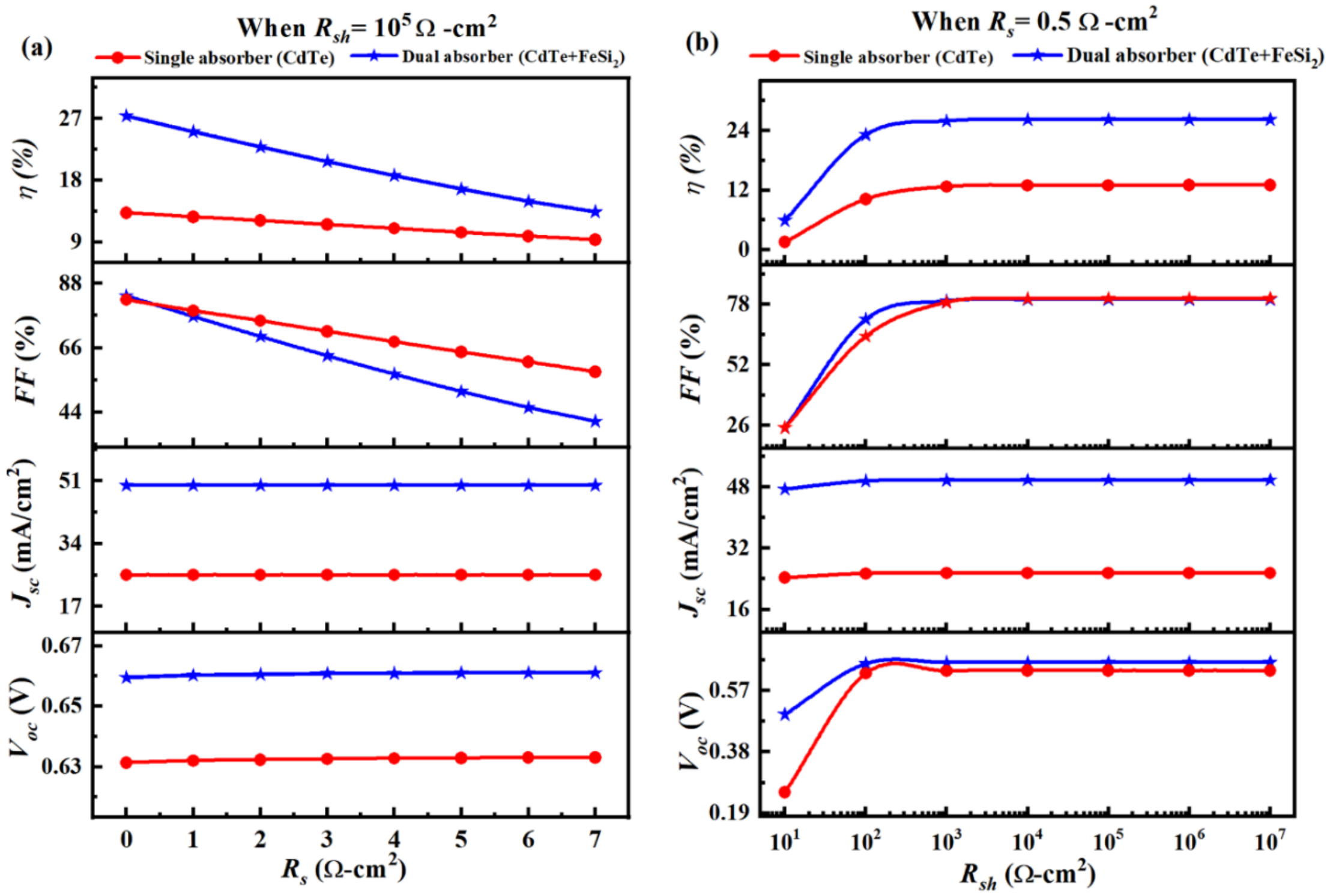

**Figure 6.** The effect of varying (a) series $R_s$, and (b) shunt $R_{sh}$ with a single and dual absorber.

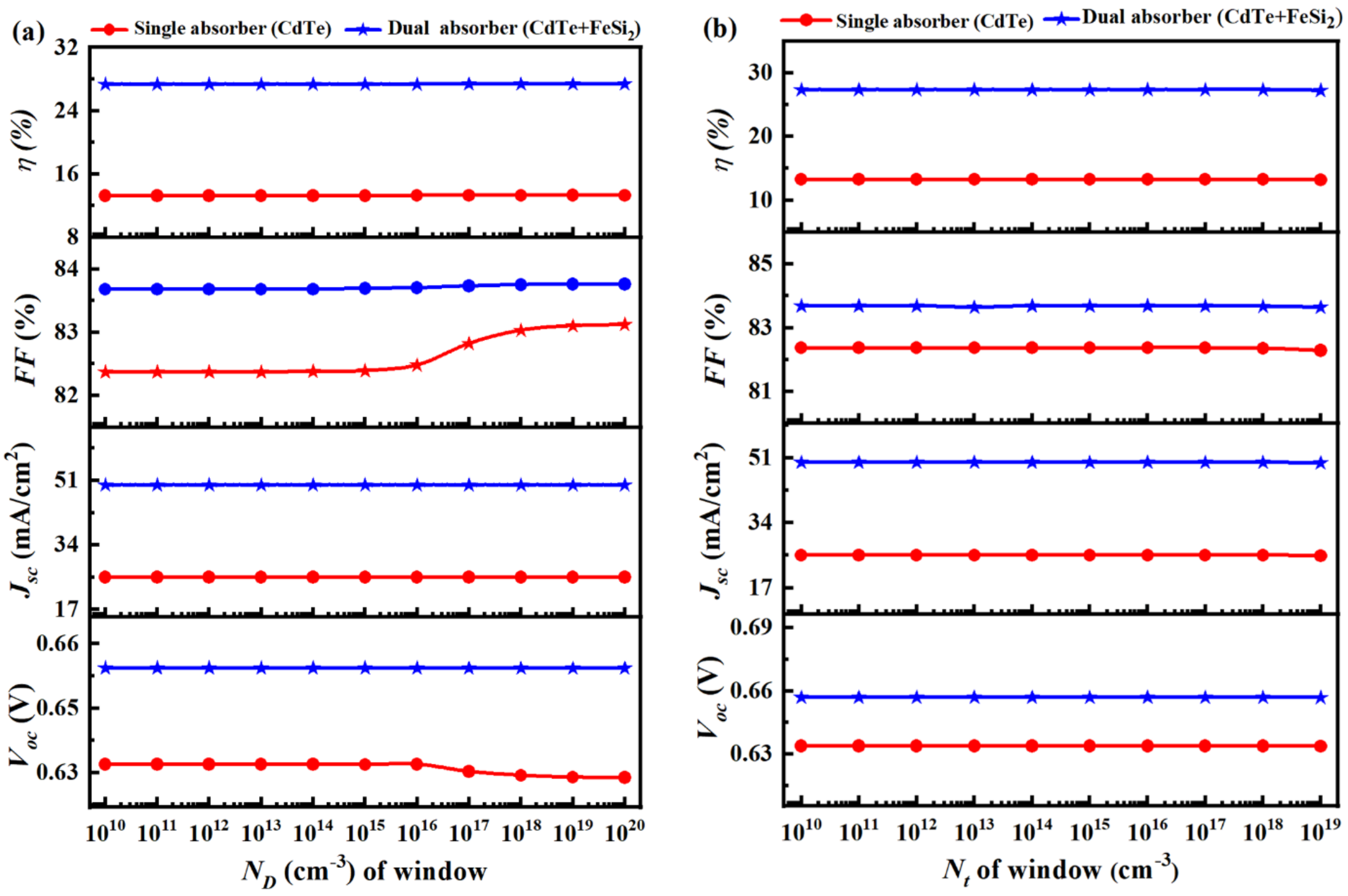

**Figure 7.** The effect of varying (a) $N_D$, and (b) $N_t$ with a single and dual absorber.

### 3.7 Impact of defect density $In_2S_3$ window layer

Studying in the window layer ($In_2S_3$), the $N_t$ on the PV performance of the solar cell, the density of neutral type acceptor is varied from $10^{10}$ to $10^{19}$ $cm^{-3}$ while maintaining the $N_t$ in solar cell performance parameters. **Figure 7(b)** displays that the values of $V_{oc}$ remained constant but after some time slightly decreases and in the case of $J_{sc}$ were also the same. By thinking about low cost and growing percentage of fabrication, the optimized value searched of the $N_t$ in $In_2S_3$ material for this proposed cell is found to be $10^{15}$ $cm^{-3}$ and which results in 27.35%.

### 3.8 Impact of interface defect density

**Figure 8** depicts how the density of interface defects ($N_{it}$) influences cell performance. The likelihood of detecting carriers at the interface increases as the interface state density ($n_t$) between CdTe and $FeSi_2$, CdTe, and $In_2S_3$ layers increases. As a result, carriers are more likely to be grabbed by their regional counterparts, resulting in low $J_{sc}$ and a reduction in $\eta$. To investigate the effect of $n_t$ on $J$–$V$ characteristics, the simulations at $n_t$ between various active layers were adjusted from a range of $1 \times 10^{10}$ $cm^{-2}$ to $1\times 10^{19}$ $cm^{-2}$. As seen in **Figure 8**, the $\eta$ of the cell decreases linearly as the $N_{it}$ increases. The projected solar cell's reverse saturation current rises as the $J_{SC}$ falls. For the suggested design, it is established that $10^{10}$ $cm^{-3}$ is the ideal value chosen for CdTe/$FeSi_2$ and CdTe/$In_2S_3$ interfaces. This value is perfect when taking into account the cheap cost and manufacturing process growth rate.

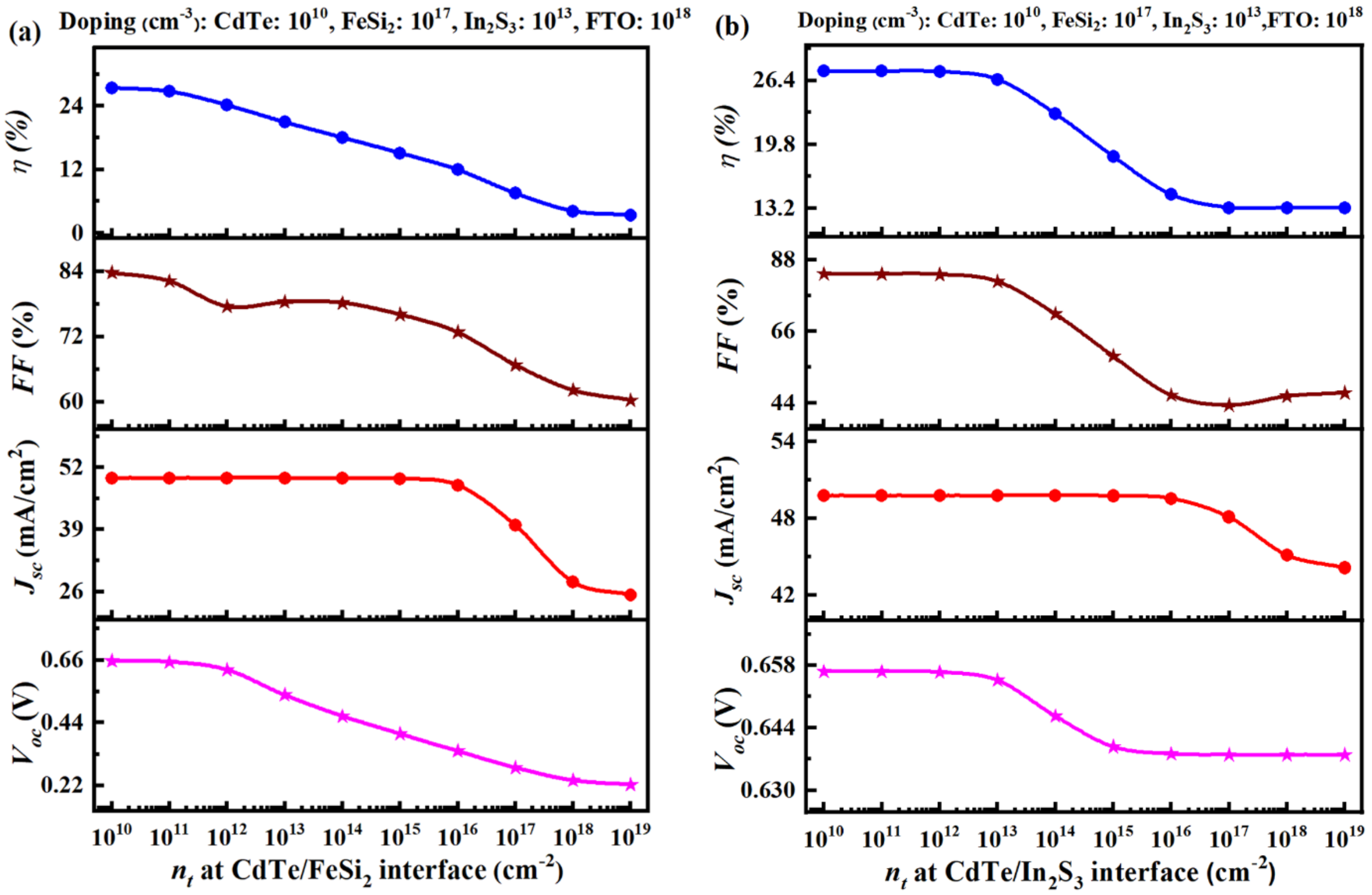

**Figure 8.** The impact $n_t$ at (a) CdTe/$FeSi_2$ interface and (b) CdTe/$In_2S_3$ interface.

### 3.9 The influence of the cell operating temperature

The functionality of CdTe solar cells with single and dual absorbers is tested when the temperature ($T$) of the cell varies from 275 to 475 K, the remaining parameters are maintained as mentioned in **Table 1**. **Figure 9** displays the impact of variations of the temperature on $J_{sc}$, $FF$, $V_{oc}$, and the $\eta$ for all the cases with single (CdTe) and dual (CdTe+$Fesi_2$) absorbers. Solar cell properties are strongly dependent on temperature since $V_{oc}$, $FF$, $J_{sc}$, and $\eta$ are temperature-dependent parameters. It should be observed that $V_{oc}$ fluctuates inversely proportionately with $T$, since the saturation current decreases fast with rising temperature, while the opposing saturation current increases. Here $V_{oc}$ is a current-dependent saturation parameter. During temperature rise, the $J_{sc}$ remains nearly constant. The suggested solar cell structure responds well to temperature changes. For the low value of temperatures, the $\eta$ is high enough. At 275 K, the $\eta$ is 29.48%, but at 475 K it decreases to 12.7%. For taking higher temperatures, $\eta$ decreases. The value of $\eta$ is low which is observed during high temperatures as a result of the temperature requirement parameters such as $\mu_h$, $\mu_n$, $E_g$, and carrier concentration of materials. $E_g$ values become unbalanced as temperature rises, and this reduces the conversion efficiency in both the electron-hole recombination in the unstable $E_g$. Single (CdTe) and with a dual (CdTe+$FeSi_2$) absorber, the remaining $J_{sc}$ is constant. With $FeSi_2$ as the absorber layer, $J_{sc}$ is always higher than conventional structures across 24.2 mA/$cm^2$ for all temperature variations. From **Figure 9** it is clear that both $V_{OC}$ and cell $\eta$ are reduced linearly in the case of basic CdTe cells. However, after incorporating the $FeSi_2$ absorber layer, these two characteristics are lowered less linearly, demonstrating the all-encompassing benefit of using the $FeSi_2$ absorber layer in a CdTe-based solar cell.

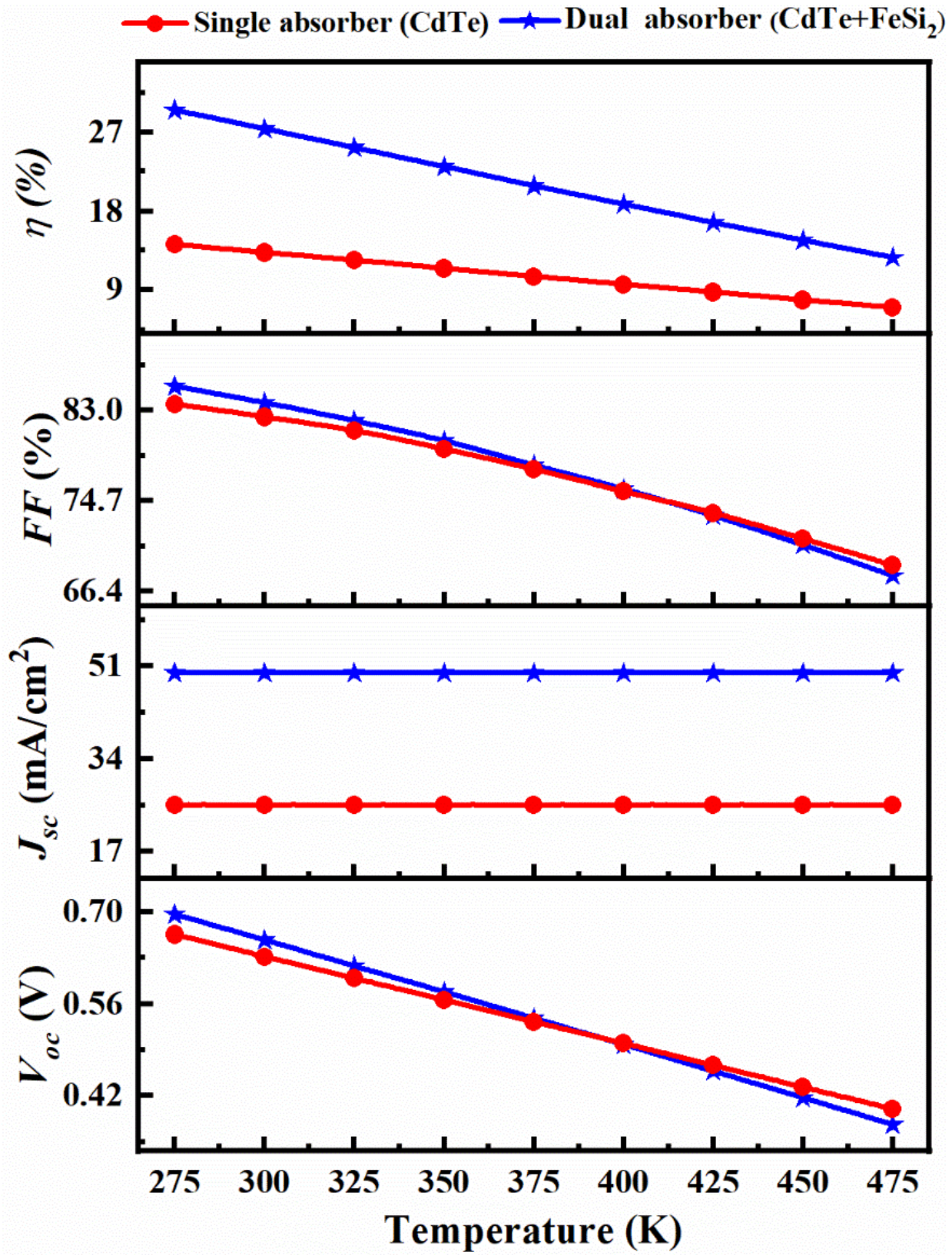

**Figure 9.** The temperature dependency of the cell's PV characteristics.

### 3.10 *J-V* properties of CdTe-based solar cells

The *J-V* characteristics of CdTe- based solar cells are shown in **Figure 10**. **Table 2** represents the summary of PV properties of the suggested solar cell structure with a 1 µm thin (CdTe+$FeSi_2$) absorber layer, which can be found to be greater than the predicted performance of a CdTe-based solar cell with a 0.5 µm thin absorber. However, the *FF* has almost increased by 1.31% in recommended cells (including a 0.5µm thin $FeSi_2$ absorber) compared to conventional cells (0.5 µm-thin CdTe absorber). For this consideration, $\eta$ is increased by about 14 %. It is clear from **Table 2** that the TFSC cell suggested herein might be more economically flexible than the other CdTe-based designs solar cells shown in **Table 2** even with just a thin, 1 µm (CdTe+$FeSi_2$) absorber layer. The arrangement also offers a way to lower the price of the absorber material used in (CdTe+$FeSi_2$) based solar cells. As a result, the suggested solar cell outperforms previous CdTe and $FeSi_2$-based solar cell structures.

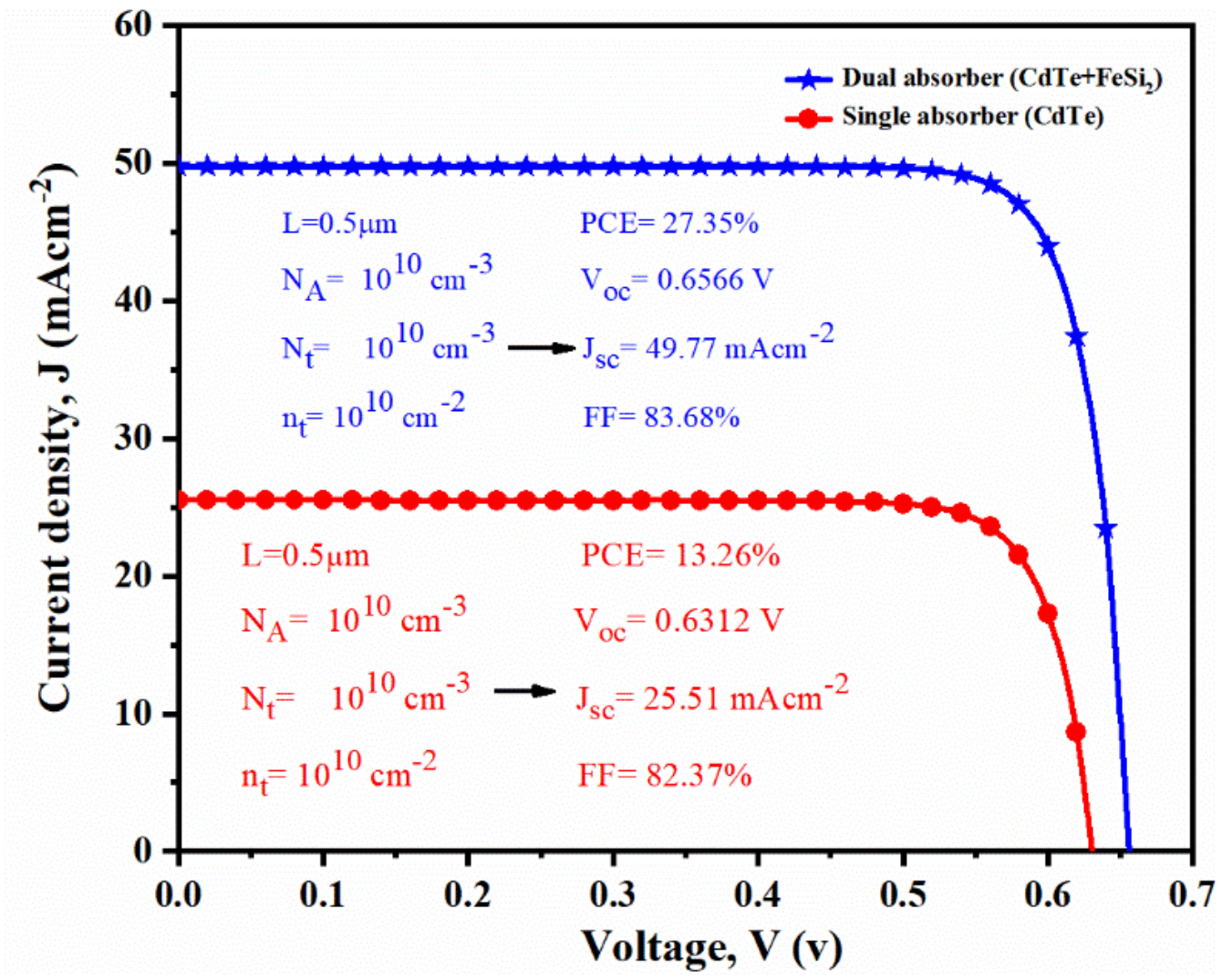

**Figure 10.** The *J-V* characteristics of the CdTe-based solar cells.

**Table 2.** A summary of the reported theoretical and experimental works.

| Type | Cell structure | Cell thickness (µm) | $V_{oc}$ (V) | $J_{sc}$ (mA/cm$^2$) | FF (%) | $\eta$ (%) | Ref. |
|---|---|---|---|---|---|---|---|
| *E* | ITO/$SnO_2$/CdS/CdTe/graphite paste/AgSC | 0.1 µm (CdS)/ 4-5 µm (CdTe) | 0.700 | 18.50 | 52.5667 | 6.8184 | [57] |
| *E* | Glass/FTO/CdS/CdTe | 1.1 µm (CdTe) | 0.63 | 38.5 | 0.33 | 8 | [58] |
| *E* | FTO/$SnO_2$/CdS/CdTe | 0.15 µm (CdS)/ 3.5 µm (CdTe) | 0.81 | 21.8 | 67 | 11.8 | [59] |
| *E* | Glass/n-$SnO_2$/n-CdS/p-CdTe | 0.09 µm (CdS)/ 5 µm (CdTe) | 0.835 | 23.67 | 69.1 | 13.66 | [60] |
| *E* | FTO/$SnO_2$/$CdSe_xTe_{1-x}$/ CdTe | 0.15 µm ($CdSe_xTe_{1-x}$)/ 3.35 µm (CdTe) | 0.824 | 26 | 66 | 14.1 | [59] |
| *T* | CdTe/CdS/$SnO_2$ | 2/0.025/0.25 | 0.9198 | 23.153579 | 66.81 | 14.23 | [41] |
| *T* | Glass/$SnO_2$/$Zn_2SnO_4$/CdS/CdTe/ZnTe | 0.5/0.1/0.1/0.05/1/0.1 | 0.90 | 24.92 | 0.70 | 15.8 | [61] |
| *T* | CdTe/ZnS/$SnO_2$ | 2/0.025/0.25 | 0.9121 | 23.260166 | 74.84 | 15.88 | [41] |
| *T* | CdTe/ZnO/$SnO_2$ | 2/0.025/0.25 | 0.9142 | 23.303926 | 76.37 | 16.27 | [41] |
| *T* | i-ZnO/CdS/CdTe/ZnTe | 0.1/0.05/1/0.1 | 0.90 | 26.42 | 0.782 | 16.9 | [62] |
| *T* | Glass/$SnO_2$/$Zn_2SnO_4$/CdS/CdTe/$Sb_2Te_3$ | 0.5/0.1/0.1/0.05/1/0.1 | 0.91 | 24.94 | 0.75 | 17.2 | [61] |
| *T* | CdTe/ZnSe/$SnO_2$ | 2/0.025/0.25 | 0.9112 | 23.484037 | 82.38 | 17.42 | [41] |
| *T* | CdTe/CdS/$SnO_2$ | 2/0.025/0.25 | 0.9113 | 23.4497335 | 81.41 | 17.43 | [41] |
| *T* | Glass/$SnO_2$/$Zn_2SnO_4$/CdS/CdTe | 0.5/0.1/0.1/0.05/1 | 0.90 | 24.60 | 0.80 | 17.8 | [61] |
| *T* | Glass/$SnO_2$/$Zn_2SnO_4$/CdS/CdTe/$As_2Te_3$ | 0.5/0.1/0.1/0.05/1/0.1 | 0.92 | 24.97 | 0.81 | 18.6 | [61] |
| *T* | ZnO/CdS/CdTe | 0.1/0.3/1 | 1.06 | 24.56 | 86.46 | 22.42 | [38] |
| *T* | ZnO/CdS/CdTe/ZnTe | 0.01/0.05/1/0.3 | 0.946 | 34.40 | 75.72 | 24.66 | [31] |
| *T* | $CdSe_xTe_{1-x}$ /CdTe | 0.7/2 | 0.923 | 31.421 | 83.985 | 24.354 | [33] |
| *T* | FTO/$TIO_2$/ZnO/CdS/CdTe/$V_2O_5$ | 0.1/0.225-0.25/0.1/1/0.125-0.15 | 0.811 | 38.51 | 80 | 25 | [42] |
| *T* | FTO/CdS/CdTe/NiO | 0.3/0.1/1/0.1 | 1.09 | 27.38 | 87.85 | 26.35 | [38] |
| *T* | Spiro-OMeTAD/IDL1/$CH_3NH_3PbI_3$/IDL2/$TiO_2$ | 0.035/0.01/0.25-0.6/0.01/0.05 | 1.63 | 21.4 | 74.86 | 26.3 | [32] |
| *T* | ITO/n-CdS/p-CdTe/$p^+$-CdSe | 0.05/0.1/1.5/0.1 | 1.15 | 30.66 | 88.57 | 31.11 | [63] |
| *T* | ITO/n-CdS/p-CdTe/$p^+$-CdSe | 0.05/0.1/1.5/0.1 | 1.05 | 49.23 | 85.71 | 44.14 | [63] |
| *T* | ZnO/CdS/CdTe/NiO | 0.1/0.1/1/0.1 | 1.09 | 29.09 | 87.84 | 28.04 | [38] |
| *T* | CdS/$FeSi_2$/$BaSi_2$ | 0.05/0.3/0.05 | 0.958 | 51 | 83 | 38.93 | [64] |
| *T* | a-Si/$FeSi_2$/Si | 0.26/0.9/0.04 | 1.982 | 13.31 | 0.762 | 19.80 | [65] |
| *T* | FTO/$In_2S_3$/CdTe | 0.04/0.05/0.5 | 0.6312 | 25.505526 | 82.37 | 13.26 | * |
| *T* | FTO/$In_2S_3$/CdTe/$FeSi_2$ | 0.04/0.05/0.5/0.5 | 0.656 | 49.776 | 83.68 | 27.35 | * |

Note: *E* = Experimental, *T.*= Theoretical, * This work

## 4 Conclusions

The SCAPS-1D simulation is used to determine the functionalities of thin-film solar cells using the CdTe absorber layer. On the optoelectronic output properties of the cell, the impacts of doping concentration, thickness, defect density, temperatures, and resistances are extensively explored. A 0.5 µm $FeSi_2$ absorber layer is introduced into the elementary CdTe solar cell structure along with the $In_2S_3$ window layer, creating an alternative structure (Cu/FTO/$In_2S_3$/CdTe/$FeSi_2$/Ni) to CdTe/CdS based solar cells. The modeled cell's PV properties are numerically computed and compared with existing similar structured solar cells' theoretical and experimental studies. Connecting a second absorber layer ($FeSi_2$) significantly enhances the voltage, current, and $\eta$, and also remarkably reduces the thickness of the absorber material. As a result, $FeSi_2$ can be considered an ideal component for other absorbers (CdTe) based solar cells. Optimized thicknesses of $In_2S_3$ window, CdTe and $FeSi_2$ absorbers, and FTO were found to be 0.05 µm, 0.04 µm, 0.5 µm, and 0.5 µm, respectively. The density of the impurity received at the CdTe level is preferred to be $1 \times 10^{10}$ $cm^{-3}$. For a specific number of defects, this framework represents an $\eta$ of 27.35% and a $V_{oc}$ is 0.6566 V, $J_{sc}$ is 49.78 $mAcm^{-2}$, and $FF$ is 83.68%. Introducing the $FeSi_2$ absorber layer, the output of this study provides design guidelines for a predictable CdTe and $FeSi_2$-based solar cell to increase its overall $\eta$ and significantly lessen the price of absorber material.

### Data availability

The raw/processed data required to reproduce these findings cannot be shared at this time as the data also forms part of an ongoing study.

### Declaration of interests

The authors declare that they have no known competing financial interests or personal relationships that could have appeared to influence the work reported in this paper.

### Acknowledgments

This research did not receive any specific grant from funding agencies in the public, commercial, or not-for-profit sectors. The authors are thankful to Marc Burgleman and his colleagues at the University of Electronics and Information Systems (ELIS), Department of Electronics and Information Systems, Belgium, for supplying the SCAPS software package, version 3.3.07.